\documentclass[aps, prl,reprint,superscriptaddress,bibnotes]{revtex4-1} 

\usepackage[T1]{fontenc}
\usepackage[utf8]{inputenc} 
\usepackage[australian]{babel}
\usepackage{svg}

\usepackage[a4paper,centering,hmargin=1.75cm,vmargin=2cm]{geometry} 
\usepackage{amsmath,amssymb,graphicx,bm,microtype} 
\usepackage[colorlinks,allcolors=blue!50!black]{hyperref} 
\usepackage[all]{hypcap}
\usepackage{cleveref,braket,siunitx}
\usepackage[version=4]{mhchem}
\usepackage{lipsum}
\usepackage{braket}

\usepackage{tikz}
\usepackage{subfiles}

\begin{document}

\title{Engineering continuous-variable entanglement in mechanical oscillators with optimal control}

\author{Maverick~J.~Millican}
\affiliation{School of Physics, University of Sydney, NSW 2006, Australia}
\affiliation{ARC Centre of Excellence for Engineered Quantum Systems, University of Sydney, NSW 2006, Australia}
\affiliation{Sydney Nano Institute, University of Sydney, NSW 2006, Australia}

\author{Vassili~G.~Matsos}
\affiliation{School of Physics, University of Sydney, NSW 2006, Australia}
\affiliation{ARC Centre of Excellence for Engineered Quantum Systems, University of Sydney, NSW 2006, Australia}

\author{Christophe~H.~Valahu}
\affiliation{School of Physics, University of Sydney, NSW 2006, Australia}
\affiliation{ARC Centre of Excellence for Engineered Quantum Systems, University of Sydney, NSW 2006, Australia}
\affiliation{Sydney Nano Institute, University of Sydney, NSW 2006, Australia}

\author{Tomas~Navickas}
\email{Current address: Q-CTRL, Sydney, NSW Australia}
\affiliation{School of Physics, University of Sydney, NSW 2006, Australia}
\affiliation{ARC Centre of Excellence for Engineered Quantum Systems, University of Sydney, NSW 2006, Australia}

\author{Liam~J.~Bond}
\affiliation{Institute of Physics, University of Amsterdam, Science Park 904, 1098 XH Amsterdam, the Netherlands}
\affiliation{QuSoft, Science Park 123, 1098 XG Amsterdam, the Netherlands}

\author{Ting~Rei~Tan}
\email{tingrei.tan@sydney.edu.au}
\affiliation{School of Physics, University of Sydney, NSW 2006, Australia}
\affiliation{ARC Centre of Excellence for Engineered Quantum Systems, University of Sydney, NSW 2006, Australia}
\affiliation{Sydney Nano Institute, University of Sydney, NSW 2006, Australia}

\begin{abstract}
We demonstrate an optimal quantum control strategy for the deterministic preparation of entangled harmonic oscillator states in trapped ions. The protocol employs dynamical phase modulation of laser-driven Jaynes-Cummings and anti-Jaynes-Cummings interactions. We prepare Two-Mode Squeezed Vacuum States (TMSS) in the motions of a trapped ion and characterize the states with phase-space tomography. We verify continuous-variable entanglement by measuring an Einstein-Podolsky-Rosen entanglement parameter of 0.0132(7), which surpasses the threshold of 0.25 for Reid's EPR criterion. We also perform a continuous-variable Bell test and find a violation of the Clauser-Horne-Shimony-Holt inequality, measuring 2.26(3), exceeding the entanglement threshold of 2. We also demonstrate the flexibility of our method by preparing a non-Gaussian entangled oscillator state -- a superposition of TMSS.
\end{abstract}

\maketitle

Quantum harmonic oscillators (QHO) are prevalent in quantum mechanical systems, having been realized in trapped ions~\cite{Wuerker1959}, nano-mechanical resonators~\cite{Regal2008, Poot2012}, superconducting cavities~\cite{Oliver2005, Gu2017}, and photonics~\cite{LIGO2013}. QHO play a critical role in quantum sensing~\cite{Giovannetti2004, Blatt2015}, simulation~\cite{Blatt2012}, and computing~\cite{Braun2005}. Entangling multiple oscillators offers further advantages. Continuous-variable (CV) entanglement, exhibited in states such as two-mode squeezed vacuum states (TMSS), features expanded capabilities for quantum metrology, quantum computing, and communication, including enhanced state teleportation~\cite{Milburn1999, Kumar2024} and sensing~\cite{Park2023, Metzner2024, Braun2005, Li2023, Cardoso2021}. Furthermore, TMSS approximate the non-physical state conceived in the Einstein-Podolsky-Rosen (EPR) paradox~\cite{EPR1935}, a thought experiment highlighting the incompatibility between quantum mechanics and local realism. 

Among various physical systems used to study CV entanglement of QHO~\cite{Ou1992, Bowen2003, Howell2004, Eberle2013, Peise2015, Fadel2018, Kunkel2018, Colciaghi2023, Leong2023, Li2023, Metzner2024}, ions confined in radio-frequency traps offer an attractive platform for investigating quantum control protocols for the creation and manipulation of CV entanglement. Each trapped ion introduces three highly harmonic mechanical oscillators. Laser cooling enables high-fidelity initialization of oscillator ground states~\cite{Wineland89}. Tuneable light-atom interactions couple the oscillators to the discrete-variable (DV) electronic levels of the ion, enabling non-Gaussian oscillator operations~\cite{Bazavan2024, Saner2024}, precision sensing~\cite{Blatt2015, McCormick2019, Valahu2024}, as well as CV logical qubit preparation~\cite{Fluhmann2019, Matsos2023} and error correction~\cite{De_Neeve_2022l}. Conveniently, the same spin-oscillator interactions can be used to map oscillator state information to electronic states for measurement~\cite{Fluhmann2020, Molmer2012, Wineland96}.

Recently, a hybrid CV-DV scheme was proposed to generate nonlinear effective spin-motion interactions in trapped-ion systems by leveraging the non-commutativity of spin operators in the control Hamiltonian~\cite{Sutherland2021}, and experimentally demonstrated to prepare non-trivial states in an oscillator~\cite{Bazavan2024,Saner2024}. This strategy offers the advantage that a broad variety of desired interactions can be synthesized by a small set of generating Hamiltonians at the cost of additional parasitic interactions. These parasitic effects cause unwanted errors that must be mitigated to allow high-fidelity operations and to scale the control strategy to multiple oscillators. 

In this work, we deploy optimal quantum control to synthesise high-quality operations that create CV entanglement between two oscillators. The key idea is to dynamically modulate the control Hamiltonian, which leads to non-commutativity in time, generating the desired entangling operations~\cite{Sutherland2021, Matsos2023, Matsos2024}. As a demonstration, we prepare TMSS and a non-Gaussian combination TMSS by applying numerically optimized phase controls on coherent laser-driven spin-oscillator interactions. We verify the CV entanglement with Reid's EPR criterion and a Bell test in the form of Clauser, Horne, Shimony, and Holt (CHSH)~\cite{CHSH1969}. 

A TMSS is defined as $\ket{\mathrm{TMSS}(r, \phi)} = \hat{S}(r, \phi) \ket{0, 0} = \sum_{n=0}^{\infty} \frac{(e^{i \phi}\tanh{r})^n}{\cosh{r}} \ket{n,n}$, with the two-mode squeezing operator $\hat{S}(r, \phi) = e^{r (e^{- i \phi} \hat{a}_1^\dagger \hat{a}_2^\dagger - e^{i \phi} \hat{a}_1 \hat{a}_2)}$, where $\hat{a}_1$ and $\hat{a}_2$ are the annihilation operators associated with two oscillators, $r$ quantifies the amount of squeezing and $\phi$ defines the correlation quadrature. The joint state $\ket{n, n}$ corresponds to two oscillators, each in the $n$th energy eigenstate. The EPR state is a TMSS in the limit of $r$ $\rightarrow$ $\infty$. In trapped ions, TMSS have been prepared using two-mode squeezing operations obtained through modulation of the trapping potential~\cite{Metzner2024} and dissipative approaches~\cite{Li2023}. 

\begin{figure}
     \centering
     \includegraphics[width=85mm]{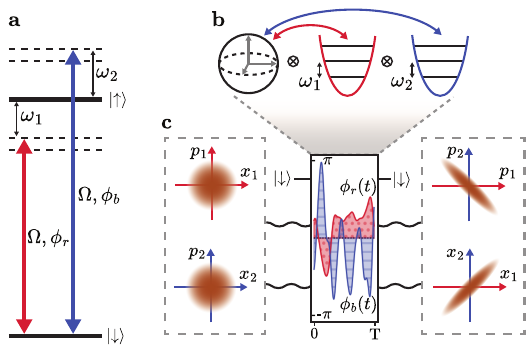}
     \label{Fig1: Overview}
     \caption{Preparation of continuous-variable entangled oscillator states using optimal quantum control. 
     \textbf{a.}~Energy diagram illustrating the two coherent interactions used by the control scheme, which include a Jaynes-Cummings (JC) interaction coupled to an oscillator with frequency $\omega_1$ (red arrow) and controllable phase $\phi_r$, and an anti-JC interaction coupled to an oscillator with frequency $\omega_2$ (blue arrow) and controllable phase $\phi_b$. Both interactions have equal coupling strength, $\Omega$.
     \textbf{b.}~Simultaneously driving the JC and anti-JC interactions couples both oscillators to a shared ancillary spin. 
     \textbf{c.}~CV entanglement is obtained by initializing the spin and oscillators to their ground state, and applying the JC-type interactions with numerically optimized phase-modulations, $\phi_r(t)$ (red function with spotted hatch) and $\phi_b(t)$ (blue function with horizontal hatch), for duration $\mathrm{T}$. Shown here is a two-mode squeezed vacuum state which exhibits non-classical position and momentum correlations between the two oscillators.
     }
\end{figure}

Here, we prepare TMSS by driving coherent time-dependent interactions that couple two different oscillators to a common spin. We use a combination of a Jaynes-Cummings (JC) and anti-JC Hamiltonian acting on oscillators 1 and 2, respectively, described by
\begin{equation}
    \label{Eq: EffHamiltonian}
        \begin{aligned}
        & \hat{H}_{\mathrm{}}(t)=\frac{\Omega}{2} \hat{\sigma}^+ \left( \hat{a}_1 e^{-i \phi_r(t)} + \hat{a}_2^{\dagger} e^{-i \phi_b(t)} \right) + \text { h.c. },
    \end{aligned}
\end{equation}
where $\Omega$ is the interaction strength and  $\hat{\sigma}^+$ is the spin-1/2 raising operator. Both $\phi_r(t)$ and $\phi_b(t)$ are tunable, time-dependent phases. 

A two-mode squeezing interaction can be obtained from the non-commutativity of $\hat{H}(t)$ in time. To better understand this, we consider the Baker-Campbell-Hausdorff (BCH) decomposition of time-sequential operators~\cite{Mielnik1970, Braun2005}, $\mathrm{e}^{ \hat{B} \delta t} \mathrm{e}^{ \hat{A} \delta t}=\mathrm{e}^{\hat{A} \delta t+ \hat{B}\delta t -\frac{1}{2}[\hat{A}, \hat{B}] \delta t^2}+\mathcal{O}\left(\delta t^3\right)$, where $\hat{A} = \hat{H}(t)$ and $\hat{B} = \hat{H}(t+\delta t)$ correspond to the Hamiltonian of Eq.~(\ref{Eq: EffHamiltonian}) at different times. Without loss of generality, we assume that $\phi_{r}(t)$ and $\phi_{b}(t)$ are piecewise constant functions of segment duration $\delta t$, where the phases at the $N$th segment are $\phi_{r_N}$ and $\phi_{b_N}$. The commutator of the time-sequential operators gives
\begin{equation} 
    \label{Eq:BCHterms}
    \begin{aligned}
        \left[\hat{A},\hat{B} \right]= &\frac{\Omega^2}{4} \hat{\sigma}_{z}\left( \tilde{\Theta}_{(r, b)} \hat{a}_1^{\dagger}\hat{a}_2^{\dagger} - \tilde{\Theta}_{(b, r)}\hat{a}_1 \hat{a}_2\right)+\\
        &\frac{\Omega^2}{2} i \sin{(\Delta_r)}\left(\hat{\sigma}_z \hat{a}_1^{\dagger} \hat{a}_1 + \ket{\uparrow}\bra{\uparrow}\right) + \\
        & \frac{\Omega^2}{2} i \sin{(\Delta_b)}\left( \hat{\sigma}_z\hat{a}_2^{\dagger} \hat{a}_2 - \ket{\downarrow}\bra{\downarrow}\right),
    \end{aligned}
\end{equation}
where $\tilde{\Theta}_{(j,k)} = e^{i(\phi_{j_{N+1}}-\phi_{k_N})}-e^{-i(\phi_{k_{N+1}}-\phi_{j_N})}$ and $\Delta_j = \phi_{j_{N+1}} - \phi_{j_{N}}$. The first term of the commutator in Eq.~(\ref{Eq:BCHterms}) corresponds to the desired spin-dependent two-mode squeezing interaction, while the second and third terms are unwanted interactions that we mitigate using optimal control. 

\begin{figure}
    \label{Fig2: Reconstructions}
     \centering
     \includegraphics[]{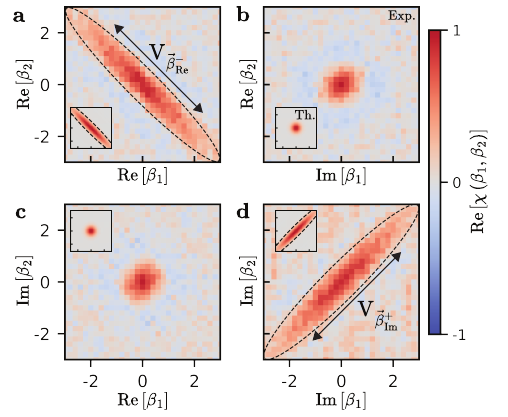}
     \caption{Experimental characteristic function tomography of a two-mode squeezed vacuum state (TMSS) with target squeezing parameter, \( r = 1 \). All panels show the real component of the joint characteristic function, \(\mathrm{Re}\left[\chi(\beta_1, \beta_2)\right]\), along four pairs of quadratures: \textbf{a.}~$\{\mathrm{Re}[\beta_1], \mathrm{Re}[\beta_2]\}$, \textbf{b.}~$\{\mathrm{Im}[\beta_1], \mathrm{Re}[\beta_2]\}$, \textbf{c.}~$\{\mathrm{Re}[\beta_1], \mathrm{Im}[\beta_2]\}$, \textbf{d.}~$\{\mathrm{Im}[\beta_1], \mathrm{Im}[\beta_2]\}$. Panels \textbf{a.} and \textbf{d.} exhibit correlation consistent with two-mode squeezing. Dashed lines plot the Gaussian functions that are fitted to the data, from which the variances \( V_{\vec{\beta}_{\mathrm{Re}}^-} \) and \( V_{\vec{\beta}_{\mathrm{Im}}^+} \) are extracted and used to quantify entanglement with Reid's EPR criterion. Panels \textbf{b.} and \textbf{c.} show negligible correlation, consistent with uncorrelated orthogonal quadratures. All panels are measured for $\beta_1>0$ and $\beta_2>0$, and we use the Hermitian property of the characteristic function, $\chi(\beta_1, \beta_2)^* = \chi(-\beta_1, -\beta_2)$, to determine $\chi(\beta_1, \beta_2)$ for $\beta_1 < 0$  and  $\beta_2 < 0 $~\cite{Valahu2023}. Insets show the theoretical characteristic function obtained from numerical simulations, and show general agreement with experiments.
     }
\end{figure}

\begin{figure*}[t]
     \centering
     \includegraphics[]{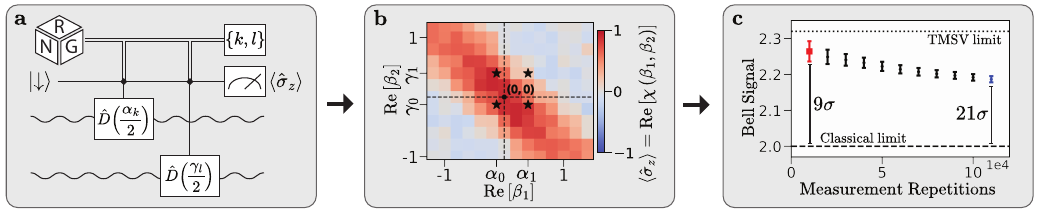}
     \label{Fig3: Bell Test}
     \caption{Continuous-variable Bell test with a TMSS. 
     \textbf{a.}~The quantum circuit used to perform the Bell test. Four predefined measurement settings $\{\alpha_k, \gamma_l\}$ with $k, l \in \{0, 1\}$ are randomly selected at each iteration of the experiment. These settings determine which displacements are performed by two subsequent SDF pulses, and the measurement is completed by a $\hat{\sigma}_z$ spin projection. This measurement gives a binary outcome, $c_m(\alpha_k, \gamma_l)$, which averaged over many repetitions corresponds to the correlation measurement, $C(\alpha_k, \gamma_l)$.
     \textbf{b.}~The correlation measurements correspond to values of the joint characteristic function, which are plotted here for a TMSS with $r=1$ in the $\{\mathrm{Re}[\beta_1], \mathrm{Re}[\beta_2]\}$ quadrature. The measurement parameters $\alpha_k$ and $\gamma_l$ are numerically optimized to maximize the Bell signal, $\mathrm{B}$. The optimal measurement parameters for a state with $r=1$ are indicated by black stars and correspond to $\{ \alpha_0, \alpha_1 \} = \{ \gamma_0, \gamma_1 \} = \{-0.1248, 0.4041\}$. 
     \textbf{c.}~The Bell signal, $\mathrm{B}$, of Eq.~(\ref{Eq:BellTest}) is plotted with respect to the total number of measurements, $M$. Values of $\mathrm{B}$ that surpass the classical limit of $2$ (dashed line) violate the CHSH inequality and indicate non-classical correlations. The dotted line at $\mathrm{B}=2.32$ bounds the CHSH inequality for a TMSS in the limit of infinite squeezing. The largest experimentally measured Bell signal with $r=1$ is $\mathrm{B}=2.26(3)$ (red square), while the most statistically significant violation of the CHSH inequality is obtained for $B=2.186(9)$ (blue dot), over 21 $\sigma$ from the classical limit. Error bars correspond to the standard deviation of the measured Bell signal determined by quantum projection noise.}
\end{figure*}

We experimentally investigate the optimal control strategy with a trapped $^{171}$Yb$^+$ ion. The harmonic oscillator frequencies along the radial directions are $\{\omega_1, \omega_2\} / 2\pi = \{ \SI{1.33}{MHz}, \SI{1.51}{MHz}\}$. Two hyperfine levels in the electronic ground states of the ion serve as an ancillary spin to facilitate interactions and measurements, experimental details provided in Supplemental Material (SM). 

The Hamiltonian of Eq.~(\ref{Eq: EffHamiltonian}) is experimentally obtained by simultaneously driving a resonant JC interaction on the first oscillator, $\hat{H}_{r_1}(t) = \frac{\Omega}{2} \left(\hat{\sigma}^+\hat{a}_1 e^{- i \phi_r(t)}\right)+ \mathrm{h.c.}$, and a resonant anti-JC interaction on the second oscillator, $\hat{H}_{b_2}(t) = \frac{\Omega}{2} \left( \hat{\sigma}^+\hat{a}_2^\dagger e^{- i \phi_b(t)}\right)+ \mathrm{h.c.}$, such that $\hat{H}_{r_1}(t) + \hat{H}_{b_2}(t) = \hat{H}(t)$, see Fig.~\ref{Fig1: Overview}. The experimental interaction strength is nominally $\Omega/2\pi = \SI{2}{kHz}$. We obtain $\phi_r(t)$ and $\phi_b(t)$ from a numerical optimizer that maximizes the fidelity of preparing a TMSS from vacuum by applying $\hat{H}(t)$, see SM. All waveforms used to generate TMSS have a predicted state overlap fidelity greater than $99.9\%$.

The preparation pulse sequence is as follows. The oscillators and ancillary spin are first initialized to their ground state, $\ket{\psi(t=0)} = \ket{\downarrow} \otimes \ket{0,0}$. We then apply $\hat{H}(t)$ for a duration $\mathrm{T}$ with optimized phase modulations, giving the target state $\ket{\psi(t=\mathrm{T})} = \ket{\downarrow} \otimes \ket{\mathrm{TMSS(r, \phi=0)}}$. Waveforms used in this work were designed with $N_{\mathrm{opt}} = 30$ optimizable segments that were filtered and resampled to higher-resolution waveforms with $N_{\mathrm{seg}} = 240$. A mid-circuit spin measurement follows to remove unwanted spin-motion entanglement that arises from experimental imperfections. The experiment proceeds if $\ket{\downarrow}$ is recorded, as a measurement of $\ket{\uparrow}$ indicates incorrect state preparation. 

We characterize the prepared TMSS through phase-space tomography. We measure the joint characteristic ($\chi$) function, $\chi(\beta_1, \beta_2) = \langle \hat{D}(\beta_1)\hat{D}(\beta_2) \rangle$, by mapping information from the harmonic oscillators to the ancillary spin using state-dependent forces (SDF)~\cite{Fluhmann2020,Valahu2023}. With this protocol, projecting the spin in the $\hat{\sigma}_z$ basis measures a point of the real part of the joint $\chi$-function, $ \langle\hat{\sigma}_z\rangle= \mathrm{Re}\left[\chi(\beta_1, \beta_2)\right] $. The real component is sufficient to characterize TMSS, as the $\chi$-functions are entirely real-valued. 

Figure~\ref{Fig2: Reconstructions} shows phase-space tomography results from joint $\chi$-function measurements for a TMSS with a target squeezing $r=1$. We find general agreement between the experiment and theory, observing a strong correlation in the $\{\mathrm{Re}[\beta_1], \mathrm{Re}[\beta_2]\}$ and $\{\mathrm{Im}[\beta_1], \mathrm{Im}[\beta_2]\}$ quadratures, implying squeezing in the multi-oscillator position and momentum quadratures, $\{{x}_1, {x}_2\}$ and $\{{p}_1, {p}_2\}$. We also generate and characterize TMSS with varying amounts of target squeezing, detailed in SM. 

Next, we quantify the continuous-variable entanglement with Reid's EPR criterion~\cite{Reid1989,Ou1992,Peise2015,Leong2023}. The EPR criterion compares the CV phase-space correlations of a state against the vacuum fluctuations of two harmonic oscillators. 

The criterion can be represented by the inequality, $\frac{1}{4\mathrm{V}_{\vec{\beta}_{\mathrm{Re}}^-} \mathrm{V}_{\vec{\beta}_{\mathrm{Im}}^+}} < \frac{1}{4},$ where $\mathrm{V}_{\vec{\beta}_{\mathrm{Re}}^-} = \mathrm{Var}\left[\chi\left(\mathrm{Re}\left[\beta/\sqrt{2}\right], - \mathrm{Re}\left[\beta/\sqrt{2}\right]\right)\right]$ and $\mathrm{V}_{\vec{\beta}_{\mathrm{Im}}^+} = \mathrm{Var}\left[\chi\left(i \mathrm{Im}\left[\beta/\sqrt{2}\right], i\mathrm{Im}\left[\beta/\sqrt{2}\right]\right)\right]$.  Given an ideal TMSS with squeezing parameter, $r$, the EPR parameter is given by $\frac{1}{4\mathrm{V}_{\vec{\beta}_{\mathrm{Re}}^-} \mathrm{V}_{\vec{\beta}_{\mathrm{Im}}^+}} = \frac{1}{4}e^{-4r}$; hence the inequality is maximally satisfied in the limit $r \rightarrow \infty$. The upper bound of $1/4$ stems from the Heisenberg uncertainty principle for unentangled states, corresponding to a squeezing parameter of $r=0$.  

We extract the variances from the reconstructed $\chi$-functions along the correlated quadratures by fitting the measurements to a two-dimensional Gaussian function (dashed lines in Fig.~\ref{Fig2: Reconstructions}a and \ref{Fig2: Reconstructions}d), giving $\mathrm{V}_{\vec{\beta}_{\mathrm{Re}}^-} = 4.4(2)$ and $\mathrm{V}_{\vec{\beta}_{\mathrm{Im}}^+} = 4.3(1)$, corresponding to $\SI{6.4(2)}{dB}$ and $\SI{6.3(1)}{dB}$ of squeezing, respectively. 
This gives a squeezing parameter $r\approx0.74$, which is less than the target $r = 1$ $(\SI{8.7}{dB})$. 
The calculated Reid criterion is $1/(4 \mathrm{V}_{\vec{\beta}_{\mathrm{Re}}^-} \mathrm{V}_{\vec{\beta}_{\mathrm{Im}}^+}) = 0.0132(7)$, which is greater than the theoretical EPR criterion of 0.0046. Discrepancies between measured and target squeezing are attributed to harmonic oscillator dephasing during the preparation of the TMSS. Numerical simulations suggest our results can be explained by dephasing that is correlated across the two oscillators at nominal dephasing rates in the range $\Gamma_{\mathrm{deph}} \approx 20-25~\mathrm{Hz}$ following methods outlined in \cite{Matsos2024}. These rates agree with independent calibrations.

We next characterize the entanglement with a Clauser, Horne, Shimony, and Holt (CHSH) Bell test. Originally designed for discrete-variable systems, the CHSH test was adapted for continuous-variable systems~\cite{Banaszek1998, Banaszek1999, Jeong2003, Milman2005}, and experimentally investigated in photonic systems~\cite{Thearle2018} and superconducting cavities~\cite{Wang2016}. 

Our experiment measures a Bell signal, B, that is classically bounded by the CHSH inequality, $\mathrm{B} \leq 2$~\cite{Bell1964, CHSH1969}. The Bell signal is constructed from four correlation measurements,
\begin{equation}\label{Eq:BellTest}
 \mathrm{B} = |C(\alpha_0,\gamma_0) + C(\alpha_0,\gamma_1) + C(\alpha_1,\gamma_0) - C(\alpha_1,\gamma_1)|,
\end{equation}
where $C(\alpha_k, \gamma_l) \in [-1, 1]$ with $k,l \in \{0, 1\}$, and measurements parameters $\alpha_k$ and $\gamma_l$ can be freely chosen. Here, the correlation measurement, $C$, is the real part of the joint $\chi$-function obtained from the expectation value, $\langle\hat\sigma_z\rangle$, of the ancillary spin, i.e., $C(\beta_1, \beta_2) = \mathrm{Re}[ \chi(\beta_1, \beta_2)] = \langle\hat\sigma_z\rangle$. Each correlation measurement is parameterized by continuous variables, $\beta_1$ and $\beta_2$, acting on oscillators 1 and 2, respectively. 

The experimental pulse sequence to perform the CV Bell test is illustrated in Fig.~\ref{Fig3: Bell Test}a. We prepare a TMSS with target squeezing parameter $r=1$ using the same control waveform as for Fig.~\ref{Fig2: Reconstructions}. Correlation measurements are then obtained by sampling four predefined points of the joint $\chi$-function, parameterized by $\alpha_k$ and $\gamma_l$, that are randomly selected at each iteration of the experiment. Choices of $\alpha_k$ and $\gamma_l$ were numerically optimized to maximize the measured Bell signal for the target TMSS, see Fig.~\ref{Fig3: Bell Test}b.

The measured Bell signal, $\mathrm{B}$, is plotted in Fig.~\ref{Fig3: Bell Test}c as a function of total measurement repetitions, $M$. The Bell signal exceeds the classical limit of 2 (dashed line), violating the CHSH inequality. We observe that $\mathrm{B}$ decreases with more measurements, which we attribute to drifts in experimental parameters that affect state preparation and measurement. The largest Bell signal is obtained at $M=10\,000$ with $\mathrm{B} = 2.26(3)$ (red square), which is near the theoretical value of 2.31 for a TMSS with $r=1$, and approaches the theoretical limit of 2.32 for any TMSS (dotted line)~\cite{Jeong2003}. The violation of the CHSH inequality is the most statistically significant at $M=110\,000$ with $B = 2.186(9)$ (blue circle), which is over $21~\sigma$ from the classical limit. This experiment takes approximately 60 minutes.

We note that our experiment uses a single ion, which does not provide spatial separation or independent readout of the two oscillators; therefore, this Bell test measurement should be considered as an entanglement witness rather than a test of nonlocality. The entanglement witnesses measured here surpass previous values from matter-based experiments~\cite{Leong2023, Wang2016}, yet fall short of those demonstrated in photonic systems~\cite{Eberle2013, Thearle2018}.

The control technique demonstrated above can be extended to create more complex entangled oscillator states. As a demonstration, we prepare a superposition of two TMSS with equal amplitudes and a relative phase shift of $\pi$, i.e., $\ket{\Psi_\times} = \frac{1}{\sqrt{2}}\left(\ket{\mathrm{TMSS}(r, 0)} + \ket{\mathrm{TMSS}(r, \pi)}\right)$. These states are non-Gaussian with potential applications as phase-insensitive electric field sensors due to their displacement sensitivity below the standard quantum limit~\cite{Cardoso2021}. 

\begin{figure}[t!]
     \centering
     \includegraphics[width=8.4cm]{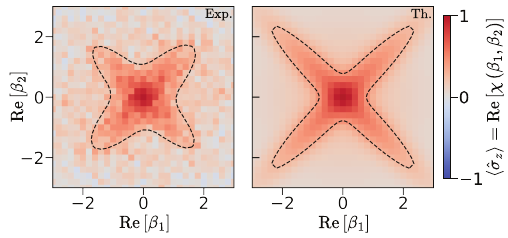}
     \caption{Joint $\chi$-function tomography of a superposition of TMSS. (Left) The experimentally prepared superposition state. (Right) numerically simulated characteristic function of the ideal target state, $\ket{\Psi_\times}$ with squeezing parameter $r=1$. The experimentally reconstructed characteristic function, $\mathrm{Re}\left[\chi(\mathrm{Re}[\beta_1], \mathrm{Re}[\beta_2])\right]$, shows features of the superposition state, with correlation along both axes, $(\mathrm{Re}[\beta_1] \pm \mathrm{Re}[\beta_2])$. 
     Dashed lines are the fits using Eq.~(\ref{Eq:Superposition State2}), from which we find squeezing parameters fitted to be $\{0.60(3), 0.69(4)\}$ along two orthogonal directions.
     }
     \label{Fig5: Superposition}
\end{figure}

We prepare $\ket{\Psi_\times}$ with $r=1$, and verify the correlations in the $\{\mathrm{Re}[\beta_1], \mathrm{Re}[\beta_2]\}$ quadratures. Figure~\ref{Fig5: Superposition} shows the experimentally reconstructed $\chi$-function, which agrees with the phase-space distribution of the ideal target state. We further characterize the prepared state by fitting a model of the theoretical $\chi$-function, which is a sum of two-dimensional Gaussian functions, 
\begin{equation}
     \begin{aligned} \label{Eq:Superposition State2}
     \mathrm{Re}\left[ \chi(\beta_1, \beta_2) \right] =\ &c_1e^{-\left(\frac{(\beta_1 + \beta_2) ^ 2}{ 2 e^{2r_1}}+\frac{(\beta_1 -\beta_2) ^ 2}{2 e^{-2r_1}}\right)}\\
     &+ c_2 e^{-\left(\frac{(\beta_1 + \beta_2) ^ 2}{ 2 e^{-2r_2}}+\frac{(\beta_1 -\beta_2) ^ 2}{2 e^{2r_2}}\right)}. 
     \end{aligned}
\end{equation}
The relative amplitudes are constrained such that $c_1 + c_2 \leq 1$, and the resulting fit gives $\{c_1, c_2\} = \{.57, .43\}$. The extracted squeezing parameters are $\{r_1, r_2\} = \{0.60(3), 0.69(4)\}$, corresponding to squeezing of $5.2(2) $ and $6.0(3)~\mathrm{dB}$ along the axes $(\mathrm{Re}[\beta_1] - \mathrm{Re}[\beta_2])$ and $(\mathrm{Re}[\beta_1]+ \mathrm{Re}[\beta_2])$, respectively. The experimentally fitted squeezing parameters are smaller than the target of $r=1$. Numerical simulations suggest that this discrepancy is dominated by correlated oscillator dephasing during state preparation, consistent with imperfect preparation of TMSS.

More broadly, the optimal control framework could be extended to synthesize a wider range of interactions. Combinations of different JC-type Hamiltonians have been shown to provide universal control over oscillator systems~\cite{Lloyd1999, Braun2005, Sutherland2021}. Time-domain dynamical modulations of combinations of JC-type Hamiltonians provide an added toolset to synthesize desired interactions. Single-mode (tri-)squeezing and beam splitter interactions also emerge naturally from the BCH time-sequential commutator used in this work~\cite{Sutherland2021, Bazavan2024}. Furthermore, modulation through optimal control can mitigate unwanted parasitic terms, allowing high-fidelity preparations of a broad class of Gaussian and non-Gaussian entangled oscillator states. 

This framework also allows for additional JC-type interactions to be included, enabling the creation of multi-partite (>2) entangled oscillator states. Our protocol suffers from classical computational difficulty in the numerical simulations to find optimal controls. This challenge can be potentially overcome by using basis sets tailored for the generating Hamiltonians and target states, e.g., the basis of Gaussian states~\cite{Walschaers2020,Bond2024,Bourassa2021,banic2025}. 

The imperfect state preparation observed across all experiments arises primarily due to oscillator dephasing during state preparation, notably for higher-energy states and longer experimental run times. Improved calibration and increased oscillator stability will be important for future experiments. Furthermore, robustness to thermal state occupation and oscillator dephasing could be explored when defining the cost function used to find optimal controls, following methods outlined in \cite{Matsos2023}.

\appendix
\section*{Supplemental Material:}

\section{Experimental Details}
Our experiment is performed with a single \ce{^{171}Yb^+} ion confined in a room-temperature ``blade-style'' Paul trap. We measure a heating rate of 0.2 quanta per second~\cite{Milne2019}, and a motional coherence time of $50$~ms~\cite{Matsos2024}, measured with a superposition between Fock states $\ket{0}$ and $\ket{1}$ in a Ramsey-type experiment. Harmonic oscillators along the radial direction of the trapping potential have frequencies $\{\omega_1, \omega_2\} / 2\pi = \{ \SI{1.33}{MHz}, \SI{1.51}{MHz}\}$. We encode an ancillary spin in the magnetically insensitive electronic states of the \ce{^2S_{1/2}} hyperfine ground state, with labels $\ket{\downarrow} = \ket{F=0, m_F=0}$ and $\ket{\uparrow} = \ket{F=1, m_F = 0}$. The frequency splitting is $\omega_0/2\pi = \SI{12.64}{GHz}$, and we measure a coherence time from a Ramsey sequence of $T_2^* = \SI{8.7}{s}$~\cite{Tan2023}. Measurements of the spin are performed in the $\hat{\sigma}_z$ basis with laser-induced state-dependent fluorescence~\cite{Olmschenk2007}. 

Coherent spin-oscillator interactions to implement the Jaynes-Cummings (JC) and anti-JC Hamiltonians are enacted by stimulated Raman transitions with frequencies equal to $\omega_0-\omega_1$ and $\omega_0+\omega_2$. A pair of orthogonal beams incident at the ion's position is obtained from a $\SI{355}{nm}$ pulsed laser. An acousto-optic modulator (AOM) in the path of one of the beams allows arbitrary phase, frequency, and amplitude modulation of these spin-oscillator interactions. 

An SDF is implemented by simultaneously driving the red and blue-sidebands of the $j$th oscillator, giving the Hamiltonian $\widehat{H}_\mathrm{SDF} = \frac{\Omega}{2}\hat{\sigma}_x ( \hat{a}_j^\dagger e^{i \phi_m} + \hat{a}_j e^{-i \phi_m})$, where $\phi_m = (\phi_b - \phi_r)/2$. Applying $\widehat{H}_\mathrm{SDF}$ results in a conditional displacement operation, $\widehat{CD}_j(\beta) = \mathrm{exp}(\hat{\sigma}_x (\beta \hat{a}_j^\dagger - \beta^* \hat{a}_j ))$. The displacement parameter, $\beta = - i \frac{\Omega}{2} \tau e^{i \phi_m}$, is fully controllable by varying the pulse duration, $\tau$, and phase, $\phi_m$. Joint phase-space tomography is performed by sequentially applying SDFs on the two oscillators with displacement parameters $\beta_1/2$ and $\beta_2/2$, giving the state $\ket{\psi} = \widehat{CD}_2(\beta_2/2)\widehat{CD}_1(\beta_1/2) (\ket{\downarrow} \otimes \ket{\mathrm{TMSS}})$~\cite{Fluhmann2019, Valahu2023}.

\section{Numerical Optimization}

Optimal pulses for preparing target entangled oscillator states are generated via gradient-based numerical optimization using Q-CTRL’s graph-based optimizer Boulder Opal~\cite{boulder_opal1, boulder_opal2}, following a similar approach to Ref.~\cite{Matsos2023}. We consider the control Hamiltonian, 
\begin{equation}
    \label{Eq:EffHamiltonian Appendix}
    \begin{aligned}
    & \hat{H}_{\mathrm{}}(t)=\frac{\Omega}{2} \hat{\sigma}^+ \left( \hat{a}_1 e^{-i \phi_r(t)} + \hat{a}_2^{\dagger} e^{-i \phi_b(t)} \right) + \text { h.c. },
    \end{aligned}
\end{equation}
which corresponds to Eq.~(1) in the main text. The control phases $\phi_r(t)$ and $\phi_b(t)$ are modeled as piecewise-constant functions with optimizable segments. 

The numerical optimization minimizes a cost function, $C$, which penalizes both the state overlap infidelity and the pulse duration,
\begin{equation}
    C = 1 - \mathcal{F} + \epsilon \frac{T}{T_{\mathrm{max}}},
\end{equation}
where $\mathcal{F} = \left|\bra{\downarrow, \psi_{\mathrm{target}}} \hat{U} \ket{\downarrow, 0, 0} \right|^2$ is the simulated state overlap fidelity between the target state and the evolved state. Here, $\hat{U} = \exp\left(-i \int_0^T \hat{H}(t)\, dt\right)$ is the time evolution operator obtained from the control Hamiltonian of Eq.~(\ref{Eq:EffHamiltonian Appendix}). The total pulse duration, $T$, is optimizable and constrained by  $0 < T \leq T_{\mathrm{max}}$, where $T_\mathrm{max} = \SI{2}{ms}$; this is chosen such that $T_\mathrm{max}$ is much smaller than the motional coherence time. Parameter $\epsilon$ is heuristically chosen to be 0.05 to limit the total pulse duration while ensuring that the final state infidelity is less than $\epsilon$. All fidelities found by the optimizer for preparing TMSS in this work are higher than $99.9\%$. 

Optimizations are performed in the energy eigenbasis of the harmonic oscillators, which is truncated to limit the size of the Hilbert space for computational efficiency. Truncation of the Hilbert space was uniquely defined for each optimization, and was influenced by the energy distribution of the target state. We used state vectors of dimension $\{2, n_{\mathrm{max}}, n_{\mathrm{max}}\}$ to find optimal pulses for TMSS, where $n_{\mathrm{max}} = \{6, 19, 32, 28\}$ for target squeezing, $r = \{.25, .75, 1, 1.25\}$. The main text shows results from the optimal pulse found for $r=1$, and the Fig.\ref{Fig4:1D Scans} shows results for the remaining $r$ values. The controls to create the superposition of TMSS, $\ket{\Psi_{\times}}$, were optimized for $n_{\mathrm{max}}=7$. We empirically find that optimizer convergence occurs on the order of 20-30 minutes for the Hilbert spaces used in this work. For larger Hilbert spaces ($n_{max} \approx 50$) and higher squeezing ($r=1.5$), we find that the convergence occurs on the order of hours, albeit with higher optimization failure rates.

Experimental constraints are incorporated during optimization by applying frequency filtering with a cutoff frequency chosen as $f_c \times T = 2\pi \times 10\,000$. Filtering helps prevent signal distortions when applying our controls on bandwidth-limited hardware, in particular the AOM in the path of one of the Raman beams that drives the modulated waveform. An initial waveform with $N_{\mathrm{opt}} = 30$ optimizable segments is filtered using a sinc kernel and then resampled to a higher-resolution waveform with $N_{\mathrm{seg}} = 240$. Before deployment, the control phases $\phi_r(t)$ and $\phi_b(t)$ are interpolated using cubic splines to remove discontinuities. The waveforms are generated in hardware with a Tabor Proteus P1284M arbitrary waveform generator.

\section{Tunable Squeezing}

\begin{figure*}[ht!]
     \centering
     \includegraphics[width=\linewidth]{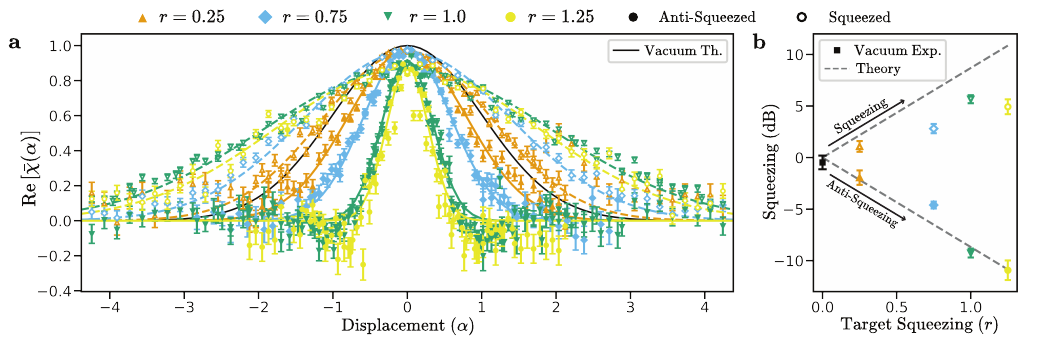}
     \label{Fig4:1D Scans}
     \caption{Experimentally measured $\chi$-function, from which we extract variances of TMSS with increasing amounts of target squeezing. 
     \textbf{a.}~Experimentally obtained $\chi$-function values averaged over both correlated quadratures; hollow markers plot reconstructions along the squeezed axis, $\mathrm{Re}\left[\bar{\chi}_\mathrm{s}(\alpha)\right]$ of Eq.~\ref{eq:chi_mean_s}; filled makers plot reconstructions along the anti-squeezed axis, $\mathrm{Re}\left[\bar{\chi}_\mathrm{as}(\alpha)\right]$ of Eq.~\ref{eq:chi_mean_as}. The measurements are repeated for TMSS with increasing target squeezing parameters $r=\{0.25, 0.75, 1.0, 1.25\}$. Solid black line plots the theoretical $\chi$-function of a vacuum state. Error bars are calculated from quantum projection noise. The reconstructed $\chi$-functions are fit to Gaussian functions, which are plotted here with (solid) dashed colored lines for fits to the reconstructions along the (anti-)squeezed axis. 
     \textbf{b.}~Variances extracted from the fits are used to quantify squeezing as a function of the target squeezing parameter, $r$. Squeezing is plotted in decibels referenced to an ideal vacuum state, with $\mathrm{V}~\mathrm{(dB)} = 10~\mathrm{log}_{10}(\mathrm{V}/\mathrm{V}_\mathrm{vac})$, where $\mathrm{V}_\mathrm{vac} = 1$ is the vacuum variance. Error bars represent three standard deviations of fitting uncertainty from the Gaussian distributions in \textbf{a.} Dashed lines plot the theoretical squeezing of ideal TMSS. The vacuum state variance observed experimentally matches a thermal occupation of $\bar{n}\approx 0.06$ in both modes. }
\end{figure*}

Here we discuss characterization of multiple TMSS states with different amounts of squeezing. 

We leverage the flexibility of our state preparation protocol to prepare TMSS with various amounts of squeezing. The numerical optimization procedure detailed above is repeated to obtain unique phase modulation waveforms for each target TMSS. For target squeezing parameters $r = \{0.25, 0.75, 1.0, 1.25\}$, the numerical optimization gives waveforms with durations $T = \{269, 650, 936, 1374\}~\mathrm{\mu s}$, based on the nominal interaction strength, $\Omega/2\pi = \SI{2}{kHz}$. Numerical optimizations return predicted state fidelities over $99.9\%$ for all pulses. 

The quality of the prepared TMSS is characterized by estimating the variances along both squeezed and anti-squeezed axes. We reconstruct 1-dimensional slices of the joint $\chi$-function in both correlated quadratures, $\{\mathrm{Re}[\beta_1], \mathrm{Re}[\beta_2]\}$ and $\{\mathrm{Im}[\beta_1], \mathrm{Im}[\beta_2]\}$. Fig.~\ref{Fig4:1D Scans}a plots the reconstructions averaged over both correlated quadratures, giving a mean measurement over the squeezed axis (hollow markers),
\begin{equation}
    \begin{aligned}
    \label{eq:chi_mean_s}
    \mathrm{Re}[\bar{\chi}_\mathrm{s}(\alpha)] = \frac{1}{2} &\mathrm{Re}\left[\chi\left(\alpha/\sqrt2, - \alpha/\sqrt2\right)\right] + \\
   \frac{1}{2} &\mathrm{Re}\left[\chi\left(i \alpha/\sqrt2, i \alpha/\sqrt2\right)\right],
\end{aligned}
\end{equation}
and the anti-squeezed axis (filled markers),
\begin{equation}
    \begin{aligned}
    \label{eq:chi_mean_as}
     \mathrm{Re}[\bar{\chi}_\mathrm{as}(\alpha)] =  \frac{1}{2} &\mathrm{Re}\left[\chi\left(\alpha/\sqrt2, \alpha/\sqrt2\right)\right] + \\
   \frac{1}{2} &\mathrm{Re}\left[\chi\left(i \alpha/\sqrt2, -i \alpha/\sqrt2\right)\right],
\end{aligned}
\end{equation}
where $\alpha \in \mathbb{R}$. The $\chi$-function follows the reciprocal variance relation outlined above which predicts larger variances of the $\chi$-function along the squeezed axes and smaller variances along the anti-squeezed axes as $r$ increases. Our experiment matches these predictions.  

We extract variances $\mathrm{Var}[\mathrm{Re}[\bar{\chi}_\mathrm{s}(\alpha)]]$ and $\mathrm{Var}[\mathrm{Re}[\bar{\chi}_\mathrm{as}(\alpha)]]$ by fitting the reconstructions of Fig.~\ref{Fig4:1D Scans}a to one-dimensional Gaussian functions. The resulting squeezing values along both squeezed and anti-squeezed axes are plotted in Fig.~\ref{Fig4:1D Scans}b for increasing target squeezing, $r$. We find general agreement with the theory (dashed line), where the magnitude of both squeezing and anti-squeezing increases with target squeezing. The highest squeezing observed in this experiment was $\SI{5.7}{dB}$ for target squeezing $r=1$. 

We observe a small offset in the experimentally measured variance of a vacuum state ($r=0$, black square in Fig.~\ref{Fig4:1D Scans}b). We attribute this to imperfect ground state cooling, which gives an estimated initial thermal occupation of $\bar{n}\approx0.06$ in both oscillators. 

\section{Phase-Space Relations}

We use two reciprocal phase-space representations of the quantum harmonic oscillator to describe the mechanical motions of a trapped ion. Both the Wigner function and the $\chi$-function contain complete information about the oscillator state, and are related to one another via a Fourier transform. 

The $\chi$-function, $\chi(\beta)$, is the expectation value of the displacement operator of a harmonic oscillator, 
\begin{equation}\label{Eq: Characteristic (op)}
     \chi(\beta) = \langle\psi_{\mathrm{osc}}|\widehat{D}(\beta)|\psi_{\mathrm{osc}}\rangle,
\end{equation}
where $\widehat{D}(\beta) = e^{\beta \hat{a}^\dagger - \beta^*\hat{a}}$. The displacement parameter, $\beta$, is complex valued, which naturally defines orthogonal axes of the $\chi$-function, $\{\mathrm{Re}[\beta], \mathrm{Im}[\beta]\}$. 

The Wigner function is obtained by taking the Fourier transform of the $\chi$-function, 
\begin{equation}\label{Eq: Wigner}
     W(\gamma) = \frac{1}{\pi^2}\int \chi(\beta) e^{\gamma\beta^* - \gamma^*\beta}d^2\beta,
\end{equation}
where $\gamma = x + ip$. This construction of the Wigner function relates the orthogonal axes of the $\chi$-function, $\{\mathrm{Re}[\beta], \mathrm{Im}[\beta]\}$, to the orthogonal axes of the Wigner function, $\{{x}, {p}\}$, respectively. 

In the case of squeezed vacuum states, both the $\chi$-function and Wigner function can be described by completely positive two-dimensional Gaussian distributions. This introduces useful properties of the Fourier transform when relating the phase-space distributions, particularly that the variances of the two representations are inversely related, 
\begin{equation}
    \begin{aligned}
    \label{Eq: Variance 1 mode}
     \mathrm{V}_{\hat{x}} &= \frac{1}{2 \mathrm{Var}\left[\chi(\mathrm{Re}[\beta])\right]} ,\\
     \mathrm{V}_{\hat{p}}&= \frac{1}{2 \mathrm{Var}\left[\chi(i\mathrm{Im}[\beta])\right]}.
     \end{aligned}
\end{equation}
This relationship applies to all axes defined by normalized linear combinations of the orthogonal basis vectors, for instance, 
\begin{equation} \label{Eq: General Variance Relation (1 mode)}
     \mathrm{V}_{\nu_1\hat{x}+\nu_2\hat{p}} = \frac{1}{2 \mathrm{Var}\left[\chi(\nu_1\mathrm{Re}[\beta]+i \nu_2\mathrm{Im}[\beta])\right]}, 
\end{equation}
where $\nu_1$ and $\nu_2$ are constants that define the normalized linear combination of basis vectors. 

The single-mode description above can be generalized to a multi-oscillator description by replacing the single oscillator displacement parameterized by $\beta$ with a multi-oscillator displacement on $N$ quantum harmonic oscillators, $\widehat{\tilde{D}}(\vec{\beta}) = \prod\limits_{n=1}^{N} \widehat{D}_n({\beta_n})$, parameterized by $\vec\beta$, giving,
\begin{equation}
     \chi(\vec{\beta}) = \langle\psi_{\mathrm{osc}, N}|\widehat{\tilde{D}}(\vec{\beta})|\psi_{\mathrm{osc}, N}\rangle.
\end{equation}

Squeezed vacuum states in systems of multiple oscillators have the same reciprocal variance relations as those described in Eq.~(\ref{Eq: General Variance Relation (1 mode)}) which can be used to characterize continuous-variable correlations between multiple quantum harmonic oscillators.

Here we consider the example of a TMSS with CV correlations in the $\{{x}_1, {x}_2\}$ and $\{{p}_1, {p}_2\}$ quadratures. This state exhibits reduced variance along the correlated position and momentum axes, $\mathrm{V}_{\hat{x}^{-}}=\mathrm{V}_{\left(\hat{x}_1 - \hat{x}_2\right)/\sqrt{2}}$ and $\mathrm{V}_{\hat{p}^{+}}=\mathrm{V}_{\left(\hat{p}_1 + \hat{p}_2\right)/\sqrt{2}}$. By measuring the $\chi$-function along quadratures defined by $\{\mathrm{Re}[\beta_1], \mathrm{Re}[\beta_2]\}$ and $\{\mathrm{Im}[\beta_1], \mathrm{Im}[\beta_2]\}$, we can obtain variances, $\mathrm{V}_{\vec{\beta}_{\mathrm{Re}}^-} = \mathrm{Var}\left[\chi\left(\mathrm{Re}\left[\beta/\sqrt{2}\right],- \mathrm{Re}\left[\beta/\sqrt{2}\right]\right)\right]$ and $\mathrm{V}_{\vec{\beta}_{\mathrm{Im}}^+} = \mathrm{Var}\left[\chi\left(i \mathrm{Im}\left[\beta/\sqrt{2}\right], i\mathrm{Im}\left[\beta/\sqrt{2}\right]\right)\right]$. 

Following the reciprocal relationship defined in Eq.~(\ref{Eq: General Variance Relation (1 mode)}), we obtain the relationship, 
\begin{equation}\label{Eq: Variance Relation (appendix)}
     \begin{split}
     \mathrm{V}_{\hat{x}^-} = \frac{1}{2 \mathrm{V}_{\vec{\beta}_{\mathrm{Re}}^-}}, \\ 
     \mathrm{V}_{\hat{p}^+} =\frac{1}{2 \mathrm{V}_{\vec{\beta}_{\mathrm{Im}}^+}}.\\
     \end{split}
\end{equation}
This relationship allows us to quantify the Wigner function variances from direct measurements of the $\chi$-function for squeezed vacuum states. 

Reid's EPR criterion~\cite{Reid1989} is originally defined by the inequality,
\begin{equation}
    \label{Eq: ReidCriteria}
     \mathrm{V}_{\hat{x}^-} \mathrm{V}_{\hat{p}^+} < \frac{1}{4},
\end{equation}
We use the reciprocal relationship between the Wigner and $\chi$-functions outlined above to evaluate the EPR criterion through measurements of the $\chi$-function for TMSS. The reciprocal EPR criterion takes the form, 
\begin{equation}
     \label{Eq: Reid Chi Appendix}
      \frac{1}{4\mathrm{V}_{\vec{\beta}_{\mathrm{Re}}^-} \mathrm{V}_{\vec{\beta}_{\mathrm{Im}}^+}} < \frac{1}{4},
\end{equation}
and is used throughout the main text. This new definition of Reid's EPR criterion is compatible with experimental phase-space tomography available in trapped ion systems, where the variances $\mathrm{V}_{\vec{\beta}_{\mathrm{Re}}^-}$ and $\mathrm{V}_{\vec{\beta}_{\mathrm{Im}}^+}$ are obtained through measurements of the joint $\chi$-function. 

\section{Bell Test Details}
\label{Appendix: Bell Test Details}
Here we provide details on the CHSH inequality measurement protocol. 

We evaluate a Bell signal, B, defined in terms of four correlation measurements,
\begin{equation}\label{Eq:BellTestSM}
\mathrm{B} = \left| C(\alpha_0,\gamma_0) + C(\alpha_0,\gamma_1) + C(\alpha_1,\gamma_0) - C(\alpha_1,\gamma_1) \right|,
\end{equation}
following the CHSH formulation~\cite{Bell1964, CHSH1969}. This is the same definition of the Bell signal as Eq.~(3) in the main text. 

Correlation measurements for the CHSH inequality of Eq. (\ref{Eq:BellTestSM}) are obtained by applying two sequential SDFs followed by a measurement of the shared ancilla, giving a binary measurement outcome $c_m(\alpha_k, \gamma_l) \in \{-1, 1\}$, as described in Fig.~3a of the main text. Spin measurements ensure that expectation values of dichotomic observables are used to compute $\mathrm{B}$ for the CV CHSH inequality \cite{Jeong2003}. 

The displacement parameter of each SDF at the $m$th measurement is randomly chosen with equal probability from one of the four possible measurement settings, $(\alpha_k, \gamma_l) \in \{(\alpha_0, \gamma_0), (\alpha_0, \gamma_1), (\alpha_1, \gamma_0), (\alpha_1, \gamma_1)\}$. The random sampling is performed in real-time using a permuted congruential type pseudo-random number generator, specifically the Xor-Shift High-Speed with Random Rotation (XHS-RR) algorithm, implemented on the field-programmable gate array of the experimental control hardware. 

We repeat the above procedure over $M$ trials, giving $M_{k,l}$ measurement outcomes for each measurement setting, with $M = \sum_{k,l} M_{k,l}$. Correlation measurements are then calculated from the average of the outcomes, giving the $\chi$-function parameterized by $\alpha_k$ and $\gamma_l$, 
\begin{equation}
    C(\alpha_k,\gamma_l) = \frac{1}{M_{k,l}}\sum_mc_m(\alpha_k,\gamma_l) = \chi(\alpha_k, \gamma_l).
\end{equation}
Uncertainties in the correlation measurement outcomes are calculated from quantum projection noise under the assumption that they are independent, identically distributed, and drawn from a binomial distribution, giving the variance $\mathrm{Var}\left[C(\alpha_k, \gamma_l)\right] = (1 - C(\alpha_k, \gamma_l)^2)/M_{k,l}$. The total variance of the Bell signal is then $\mathrm{Var}\left[\mathrm{B}\right] = \sum_{k,l} \mathrm{Var}\left[C(\alpha_k, \gamma_l)\right]$.

Choice of the optimal measurement parameters is informed by numerical simulation of the above protocol on the target state. For this experiment, we restrict the correlation measurements to lie in the $\{\mathrm{Re}[\beta_1], \mathrm{Re}[\beta_2]\}$ quadrature of the joint $\chi$-function, and constrain the measurement parameters to a symmetric set, namely $\alpha_{0}=\gamma_{0}$ and $\alpha_{1}=\gamma_{1}$. We then maximize the Bell signal of Eq.~(\ref{Eq:BellTestSM}) by numerically calculating the $\chi$-function of the target state, giving optimal measurement parameters $\{ \alpha_0, \alpha_1 \} = \{ \gamma_0, \gamma_1 \} = \{-0.1248, 0.4041\}$ for a TMSS of r=1, see black stars in Fig.~3b in the main text. 

\begin{acknowledgments}
We thank B. Baragiola and Z. Huang for fruitful discussion. We were supported by the Australian Research Council (FT220100359), the U.S. Office of Naval Research Global (N62909-24-1-2083), the U.S. Army Research Office Laboratory for Physical Sciences (W911NF-21-1-0003), the U.S. Air Force Office of Scientific Research (FA2386-23-1-4062), the Sydney Quantum Academy (MJM), the University of Sydney Postgraduate Award scholarship (VGM), and H.\ and A.\ Harley. We acknowledge the use of AI-assisted tools, including language models, to aid the initial drafting of the manuscript text. 
\end{acknowledgments}

\section*{Data availability}

The experimental data are available in an online repository at https://zenodo.org/records/15858807 \cite{DataAvailability}.


%

\end{document}